\documentclass[10pt, conference]{IEEEtran}
\IEEEoverridecommandlockouts

\usepackage{cite}

\ifCLASSINFOpdf

\else

\fi

\let\ssstyle=\scriptscriptstyle
\newcommand{\define}{\newcommand}
\newcommand{\bt}{\begin{tabular}}
\newcommand{\et}{\end{tabular}}
\newcommand{\be}{\begin{equation}}
\newcommand{\ee}{\end{equation}}

\newcommand{\bd}{\begin{displaymath}}
\newcommand{\ed}{\end{displaymath}}
\newcommand{\ba}{\begin{array}}
\newcommand{\ea}{\end{array}}
\newcommand{\bn}{\begin{enumerate}}
\newcommand{\en}{\end{enumerate}}
\newcommand{\bds}{\begin{description}}
\newcommand{\eds}{\end{description}}
\newcommand{\bi}{\begin{itemize}}
\newcommand{\ei}{\end{itemize}}
\newcommand{\bc}{\begin{center}}
\newcommand{\ec}{\end{center}}
\newcommand{\bqa}{\begin{eqnarray}}
\newcommand{\eqa}{\end{eqnarray}}

\define{\ul}{\noindent \underline}
\define{\Gamb}{{\bf \Gamma}}
\define{\Sigb}{{\bf \Sigma}}
\define{\Lambs}{{\bf \Lambda}_{s}}
\define{\LambR}{{\bf \Lambda}_{R}}
\define{\OmB}{{\bf \Omega}_{oB}}
\define{\Omo}{{\bf \Omega}_o}
\define{\Omb}{{\bf \Omega}}
\define{\Oms}{{\bf \Omega}_s}
\define{\alpb}{{\bf \alpha}}
\define{\alpbb}{\bar{\bf \alpha}}
\define{\betb}{{\bf \beta}}
\define{\betbb}{\bar{\bf \beta}}
\define{\tr}{\mbox{Tr}}
\define{\CR}{Cram\'{e}r-Rao~}
\define{\Mu}{{\it MUSIC}}
\define{\ES}{{\it ESPRIT}}
\define{\Mi}{{\it Min-Norm}}
\define{\SR}{{\it Subspace Rotation}}
\define{\ath}{{{\bf a}(\theta_k)}}
\define{\Ath}{{\bf A}(\theta)}
\define{\AthH}{{\bf A}^{H}(\theta)}
\define{\athd}{{{\bf a}^{(1)}(\theta_k)}}
\define{\bo}{{\bf 1}}
\define{\bz}{{\bf 0}}
\define{\ab}{{\bf a}}
\define{\bb}{{\bf b}}
\define{\cb}{{\bf c}}
\define{\cbb}{\bar{\bf c}}
\define{\db}{{\bf d}}
\define{\eb}{{\bf e}}
\define{\ebb}{\bar{\bf e}}
\define{\ebk}{{\bf e}_k}
\define{\ebbk}{\bar{\bf e}_k}
\define{\fb}{{\bf f}}
\define{\gb}{{\bf g}}
\define{\hb}{{\bf h}}
\define{\mb}{{\bf m}}
\define{\nb}{{\bf n}}
\define{\ob}{{\bf o}}
\define{\pb}{{\bf p}}
\define{\qb}{{\bf q}}
\define{\rb}{{\bf r}}
\define{\sbb}{{\bf s}}
\define{\tb}{{\bf t}}
\define{\ub}{{\bf u}}
\define{\vb}{{\bf v}}
\define{\wb}{{\bf w}}
\define{\st}{{\bf s}}
\define{\xb}{{\bf x}}
\define{\yb}{{\bf y}}
\define{\zb}{{\bf z}}
\define{\Ab}{{\bf A}}
\define{\Psib}{{\bf \Psi}}
\define{\Xib}{{\bf \Xi}}

\define\sT{{\ssstyle T}}

\define{\Au}{{\Ab^{\uparrow}}}
\define{\Af}{{\bf A}_F}
\define{\As}{{\bf A}_S}
\define{\Bb}{{\bf B}}
\define{\Cb}{{\bf C}}
\define{\Cbh}{\hat{\bf C}}
\define{\Db}{{\bf D}}
\define{\Eb}{{\bf E}}
\define{\Fb}{{\bf F}}
\define{\Gb}{{\bf G}}
\define{\Hb}{{\bf H}}
\define{\Ib}{{\bf I}}
\define{\If}{{\bf I}_F}
\define{\Is}{{\bf I}_S}
\define{\Ibd}{{\bf I}^{\downarrow}}
\define{\Ibu}{{\bf I}^{\uparrow}}
\define{\Jb}{{\bf J}}
\define{\Kb}{{\bf K}}
\define{\Lb}{{\bf L}}
\define{\Lbb}{\bar{\bf L}}
\define{\Mb}{{\bf M}}
\define{\Nb}{{\bf N}}
\define{\Ob}{{\bf O}}
\define{\Pb}{{\bf P}}
\define{\Qb}{{\bf Q}}
\define{\Rb}{{\bf R}}
\define{\Rbh}{\hat{\bf R}}
\define{\Rsd}{\Delta {\bf R}_s}
\define{\Rba}{{\bf R}_A}
\define{\Rbse}{\stackrel{\sim}{{\bf R}_{s}}}
\define{\Rbb}{\bar{\bf R}}
\define{\Sb}{{\bf S}}
\define{\Tb}{{\bf T}}
\define{\Ub}{{\bf U}}
\define{\Vb}{{\bf V}}
\define{\Vn}{{\bf V_n}}
\define{\Wb}{{\bf W}}
\define{\Xb}{{\bf X}}
\define{\Yb}{{\bf Y}}
\define{\Zb}{{\bf Z}}
\define{\thb}{{\bf \Theta}}
\define{\Yd}{\Delta {\bf Y}}
\define{\Yt}{\tilde{\bf Y}}
\define{\Ut}{\tilde{\bf U}}
\define{\Vt}{\tilde{\bf V}}
\define{\Unt}{\tilde{\bf U}_o}
\define{\unt}{\tilde{\bf u}_o}
\define{\Ust}{\tilde{\bf U}_s}
\define{\ust}{\tilde{\bf u}_s}
\define{\Wt}{\tilde{\bf W}}
\define{\Wd}{\Delta \bf {W}}
\define{\Phb}{{\bf {\Phi}}}
\define{\Omegab}{{\bf {\Omega}}}
\define{\Gammab}{{\bf {\Gamma}}}
\define{\Phib}{{\bf {\Phi}}}
\define{\Phd}{\Delta \bf {\Phi}}
\define{\Phl}{{{\bf \Phi}^{\downarrow}}}
\define{\Phu}{{{\bf \Phi}^{\uparrow}}}
\define{\Psd}{\Delta \bf {\Psi}}
\define{\rt}{\tilde{r}}
\define{\rd}{\Delta r}
\define{\td}{\Delta \theta}
\define{\tl}{\tilde{\theta}}
\define{\ut}{\tilde{\bf u}}
\define{\Und}{\Delta {\bf U}_o}
\define{\Usd}{\Delta {\bf U}_s}
\define{\Us}{{\bf U}_s}
\define{\us}{{\bf u}_s}
\define{\Uo}{{\bf U}_o}
\define{\uo}{{\bf u}_o}
\define{\Vs}{{\bf V}_s}
\define{\vs}{{\bf v}_s}
\define{\Vo}{{\bf V}_o}
\define{\vo}{{\bf v}_o}
\define{\Aa}{||{\bf \alpha}_k||^{2}}
\define{\Usx}{{\bf U}_{sx}}
\define{\Usy}{{\bf U}_{sy}}
\define{\Usxd}{\Delta {\bf U}_{sx}}
\define{\Usyd}{\Delta {\bf U}_{sy}}
\define{\Usxt}{\tilde{\bf U}_{sx}}
\define{\Usyt}{\tilde{\bf U}_{sy}}
\define{\Usz}{{\bf U}_{sz}}
\define{\Uszd}{\Delta {\bf U}_{sz}}
\define{\Uszt}{\tilde{\bf U}_{sz}}
\define{\Ft}{\tilde{\bf F}}
\define{\Fd}{\Delta {\bf F}}
\define{\At}{\tilde{\bf A}}
\define{\Bt}{\tilde{\bf B}}
\define{\Bd}{\Delta {\bf B}}
\define{\ld}{\Delta \bar{\lambda}_k}
\define{\ldd}{\Delta \lambda_k}
\define{\lt}{\tilde{\lambda}}
\define{\lb}{\bar{\lambda}}
\define{\df}{\stackrel{\rm def}{=}}
\define{\diag}{\rm diag}
\define{\prf}{\noindent \underline{Proof:}\\}
\define{\pe}{\hfill $\Box$}
\define{\doi}{\stackrel{j\neq i}{j=1}}
\define{\dok}{\stackrel{j\neq k}{j=1}}
\define{\ds}{\displaystyle}
\define{\Usu}{{\bf U}_s^{\uparrow}}
\define{\Usl}{{\bf U}_s^{\downarrow}}
\define{\Unu}{{\bf U}_o^{\uparrow}}
\define{\Unl}{{\bf U}_o^{\downarrow}}
\define{\Oo}{{\bf \Omega}_{o}}
\define{\Ot}{{\tilde{\bf \Omega}}_{o}}
\define{\Os}{{\bf \Omega}_{s}}
\define{\Ou}{{\bf O}^{\uparrow}}
\define{\Od}{{\bf O}^{\downarrow}}
\define{\thd}{\Delta \theta}
\define{\Iu}{{\bf I}^{\uparrow}}
\define{\Il}{{\bf I}^{\downarrow}}
\define{\Otd}{{\tilde{\bf O}}^{\downarrow}}
\define{\Sr}{{\bf \Sigma}_s^{{1}\over{2}}}
\define{\Sp}{{\bf \Sigma}_s^{-1}}
\define{\Ss}{{\bf \Sigma}_s^{-{{1}\over{2}}}}
\define{\nun}{\underline{\bf n}}
\define{\nh}{\bar{\bf n}}
\define{\eq}[1]{(\ref{#1})}
\define{\dB}{\delta {\bf B}}
\define{\deb}{\delta b}
\define{\bul}{\underline{\beta}}
\define{\ape}{\stackrel{\cdot}{=}}

\define{\im}{\,\mbox{Im}}
\define{\re}{\,\mbox{Re}}

\define{\rmE}{\mathrm{E}}
\define{\rme}{\mathrm{e}}
\newcommand{\Expect}[1]{\mathop{\mathbb{E}}\left[ #1 \right ]}

\let\ssstyle=\scriptscriptstyle

\def\s0{{\ssstyle 0}}

\def\sT{{\ssstyle T}}

\define{\Pib}{{\bf \Pi}}
\define{\bs}{\begin{slide}}
\define{\es}{\end{slide}}
\define{\xz}{\epsilon_{l,0}(n)}
\define{\xx}{\epsilon_{l,1}(n)}
\define{\xc}{\epsilon_{l,2}(n)}

\usepackage{amssymb,amsmath,epsfig}
\usepackage[latin5]{inputenc}

\usepackage{color}
\usepackage{graphicx}
\usepackage{mathtools}

\usepackage{acronym}

 \usepackage[usenames,dvipsnames]{pstricks}
 \usepackage{epsfig}
 \usepackage{pst-grad} 
 \usepackage{pst-plot} 
 \usepackage[space]{grffile} 
 \usepackage{etoolbox} 
\usepackage{epstopdf}
\usepackage{cite}

\begin{document}
%
\title{Sparsity-Aware Joint Frame Synchronization and Channel Estimation: Algorithm and USRP Implementation}


\author{ \IEEEauthorblockN{
\"{O}zg\"{u}r \"{O}zdemir\IEEEauthorrefmark{1},  Ridha
Hamila\IEEEauthorrefmark{2}, Naofal
Al-Dhahir\IEEEauthorrefmark{3}, and \.{I}smail~G\"{u}ven\c{c}\IEEEauthorrefmark{1}}
\IEEEauthorblockA{\IEEEauthorrefmark{1}Department of Electrical \& Computer Engineering, North Caralina State University, Raleigh, NC
\IEEEauthorblockA{\IEEEauthorrefmark{2}Department of Electrical Engineering, Qatar University, Doha, Qatar}
\IEEEauthorblockA{\IEEEauthorrefmark{3} Electrical Engineering
Department, The University of Texas at Dallas, Richardson, TX \\Email:
\{oozdemi, iguvenc\}@ncsu.edu, hamila@qu.edu.qa, aldhahir@utdallas.edu}}\thanks{This publication was made possible by NPRP grant \# NPRP 6-070-2-024 from the Qatar National Research Fund (a member of Qatar Foundation). The statements made herein are solely the responsibility of the authors.}}

\maketitle

\begin{abstract}
Conventional correlation-based frame synchronization techniques can suffer significant performance degradation over multi-path frequency-selective channels. As a remedy, in this paper we consider joint frame synchronization and channel estimation. This, however, increases the length of the resulting combined channel and its estimation becomes more challenging. On the other hand, since the combined channel is a sparse vector, sparse channel estimation methods can be applied. We propose a joint frame synchronization and channel estimation method using the orthogonal matching pursuit (OMP) algorithm which exploits the sparsity of the combined channel vector. Subsequently, the channel estimate is used to design the equalizer. Our simulation results and experimental outcomes using software defined radios show that the proposed approach improves the overall system performance in terms of the mean square error (MSE) between the transmitted and the equalized symbols compared to the conventional method.

\end{abstract}

\IEEEpeerreviewmaketitle

\section{Introduction}
In communication systems where the information symbols are transmitted in frames, it is essential to determine the frame boundary correctly to avoid performance degradation. In conventional systems, frame synchronization is performed by correlating a known training sequence with the received data  and the point which gives the highest correlation is selected as the frame boundary~\cite{massey&tcom:72}. In multi-path fading environments where the delay spread is larger than the symbol duration, the correlation peak gets widened and it becomes difficult to identify the exact frame boundary. Moreover, if the location of the dominant multi-path component happens to be somewhere other than the first tap, this location will give the highest correlation and an incorrect frame boundary will be chosen.

Equalization aims to convert the multi-path channel into a single-tap flat-fading channel which would improve the accuracy of the frame synchronization. However, the channel estimate required for equalization is often calculated using the training symbols, with the knowledge of the frame boundary itself. Therefore, a joint synchronization and channel estimation approach is desirable for simultaneously obtaining the channel estimates and the frame boundary.

In order to tackle this problem, we model the frame boundary offset as an unknown delay introduced to the channel which we call the combined channel. This combined channel is estimated and the equalizer is designed based on this estimate. The disadvantage of this approach is that the length of the channel impulse response (CIR) may increase by an amount as large as the duration of the frame length and the channel estimation becomes more complex. However, we note that introducing delay into the channel is equivalent to padding the CIR with zeros. Therefore, the number of non-zero CIR elements that needs to be estimated does not increase and the combined channel vector becomes sparse. As a result, sparse channel estimation methods can be used to reduce complexity.

Sparse channel estimation has been considered in~\cite{cotter&tc:2002} using a matching pursuit (MP) algorithm~\cite{mallat@tsp:93,cotter&visp:99} while a technique based on least mean squares (LMS) is proposed in~\cite{gui&wcnc:13}. Neither of the methods in~\cite{cotter&tc:2002,mallat@tsp:93,cotter&visp:99,gui&wcnc:13}  consider frame synchronization. Earlier work on joint frame synchronization and channel estimation includes~\cite{wen@wicom:07} for OFDM systems, \cite{marey@vtc:07} and \cite{marey@vtc:08} for CDMA systems, and~\cite{cheng&spie:16} for optical communication systems. However none of them exploit the sparsity of the combined channel vector. To the best of our knowledge, joint frame synchronization and channel estimation using sparse methods has not been studied in the literature.

In this paper, we propose a joint frame synchronization and channel estimation method using the orthogonal matching pursuit (OMP) algorithm which exploits the sparsity of the combined CIR vector. An equalizer is designed using the estimated channel. Subsequently, the mean square error (MSE) between the equalizer output and the transmitted symbols is used to illustrate the performance of the proposed method. Both simulation results and the experimental results using universal software radio peripherals (USRPs) demonstrate the performance improvements using the proposed method.

The rest of this paper is organized as follows.  Section~II
describes the system model. In Section III, the proposed joint frame synchronization and channel estimation method is explained.  An equalizer design using the channel estimate obtained in Section III is discussed in Section IV. Simulation and experimental testbed results in Section V verify the performance improvements when the proposed method is used. The paper is concluded in Section VI.

{\bf Notation}: Vectors are represented by lower-case boldface letters. Matrices are represented by upper-case boldface letters. The transpose is denoted by  $(.)^{\sT}$ and $\Expect{.}$ is the expectation operator.

\section{System Model}

\begin{figure}[!t]
\centering
\includegraphics[width=\columnwidth]{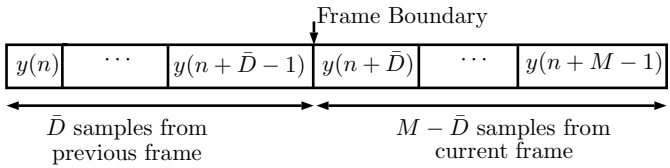}
\caption{The received samples and the illustration of frame boundary $\bar{D}$ for a data frame.}
\label{regframe:fig}
\end{figure}

We consider a frame-based communication system over a multi-path frequency-selective channel. The information symbols are transmitted in frames and the receiver must estimate the frame boundary, which is referred to as frame synchronization, and perform channel estimation. The channel estimate is used for channel equalization and demodulation to recover the information symbols at the receiver. The multi-path CIR is denoted by the vector $\hb=\left [ h_0, h_1, \cdots, h_L \right ]^{\sT}$ where $L$ is the CIR memory. If frame synchronization is performed prior to the channel estimation, then, assuming symbol-spaced sampling at the receiver, the $n$-th received symbol after frame synchronization can be written as follows
\begin{align}
 y(n)&=\sum_{l=0}^{L} x(n-l) h_l + z(n)= x(n) h_0 + \cdots +z(n),
\end{align}
where $x(n)$ is the transmitted symbol and $z(n)$ is the additive white Gaussian noise (AWGN) symbol at time $n$. Note that, in constructing $y(n)$, $x(n)$ is multiplied by $h_0$. On the other hand if frame synchronization is not performed prior to channel estimation then the received signal becomes
 \begin{align}
 \label{delayD:eq} y(n)&=\sum_{l=0}^{L} x(n-D-l) h_l + z(n),
\end{align}
where $D$ is the delay, in symbol periods, between the transmitter and the receiver. Note that, in this case $x(n-D)$ is multiplied by $h_0$. Suppose that the transmitted frames contain $M$ received samples and that we arbitrarily collect $M$ samples $\{y(n), \cdots, y(n+M-1) \}$ without the knowledge of the frame boundary denoted by $\bar{D}$. The frame boundary, $\bar{D}$, is a random number between 0 and $M-1$. If the frame boundary is equal to zero, then all of the $M$ received samples correspond to the same frame. Otherwise, some initial samples actually belong to the previous frame, as illustrated in Fig.~\ref{regframe:fig}.
The delay $D$ in~(\ref{delayD:eq}) and the frame boundary $\bar{D}$ are related by
\be
\label{barD:eq}
D=mM+\bar{D},
\ee
where $m=0,1,\cdots$ is an integer. To divide the received samples into frames, the knowledge of $m$ is not required as shown in Fig.~\ref{regframe:fig}. Therefore, we only investigate finding $\bar{D}$ for frame synchronization.

\begin{figure}[!t]
\centering
\includegraphics[width=2.3 in]{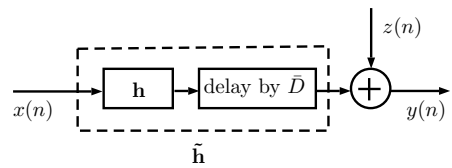}
\caption{The system model that takes the multi-path channel and delay into consideration.}
\label{sysmodel:fig}
\end{figure}

Although it is possible to perform frame synchronization and channel estimation separately, we may treat them jointly as well by defining the following delayed and zero-padded combined CIR vector $\tilde{\hb}$
\be
\label{htilde:eq}
\tilde{\hb}= [\underbrace{0, 0, \cdots, 0}_{\bar{D} \: \mathrm{zeros}} \underbrace{h_0, h_1, \cdots, h_L}_{\hb^{\sT}}, \underbrace{0, 0, \cdots, 0}_{M-\bar{D}-1 \: \mathrm{zeros}}  ]^{\sT}.
\ee
The system model that takes the multi-path channel and delay into consideration is illustrated in Fig.~\ref{sysmodel:fig}.
In~(\ref{htilde:eq}), the $M-\bar{D}-1$ zeros at the end do not have any effect as far as the input and output of the delayed and zero padded CIR $\tilde{\hb}$ is concerned. However, they ensure that the length of $\tilde{\hb}$ is $M+L$ which is independent of the value of $\bar{D}$. Note that, as long as $M\gg 1$, $\tilde{\hb}$ is a sparse vector regardless of $\hb$ being sparse or not.

\section{Joint Sparse Channel Estimation and Frame Synchronization}
\label{joint:sec}
\begin{figure}[!t]
\centering
\includegraphics[width=\columnwidth]{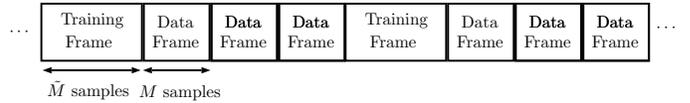}
\caption{The frame structure we assume to perform joint channel estimation and frame synchronization. One training frame is transmitted every $P$ frames. This figure assumes $P=4$. The value of $P$ is determined based on the channel coherence time. The size of the training frame is $\tilde{M}$ samples and the size of the data frame is $M$ samples.}
\label{frame:fig}
\end{figure}

In order to perform joint channel estimation and frame synchronization, we assume that a known training frame with size $\tilde{M} > M$ is transmitted periodically as shown in Fig.~\ref{frame:fig}.
The size of the training frame needs to be larger than the size of the data frame for reasons that will be explained later. The estimate of $\tilde{\hb}$ is calculated using the training frame and it is used to equalize and decode the data frames until the next $\tilde{\hb}$ estimate is calculated when the next training frame is transmitted again after $P$ frames. The period $P$ is determined according to the channel coherence time. The input-output relation of the system model illustrated in Fig.~\ref{sysmodel:fig} is
\be
\label{yn:eq}  y(n)=\sum_{l=0}^{M+L-1} x(n-l) \tilde{h}_l + z(n),
\ee
where $\tilde{h}_l$ is the $l$-th element of $\tilde{\hb}$.

Since data frames are assumed to contain $M$ samples, the receiver will collect $M$ samples as in Fig.~\ref{regframe:fig} when the current frame is a data frame. On the other hand, when the current frame is a training frame with size $\tilde{M}$, the receiver collects $\tilde{M}$ samples $\{y(n), \cdots, y(n+\tilde{M}-1) \}$.
Note that, because $\bar{D}$ is an integer in $[0, M-1]$, in the worst case, $D=M-1$ and the samples $y(n+M)$, $y(n+M+1)$, $\cdots$, $y(n+\tilde{M}-1)$ are guaranteed to be in the training frame rather than in the frame preceding the training frame. In joint channel estimation and frame synchronization a subset of these samples $y(n+\tilde{M}-N_{\mathrm{E}})$, $y(n+\tilde{M}-N_{\mathrm{E}}+1)$, $\cdots$, $y(n+\tilde{M}-1)$ are used where $N_{\mathrm{E}}$ is the number of equations. Using (\ref{yn:eq}) and the sample $y(n+\tilde{M}-1)$ the first equation is given by
\be
\label{first:eq}
y(n+\tilde{M}-1)=\sum_{l=0}^{M+L-1} x_{t,\tilde{M}-1-l} \tilde{h}_l + z(n+\tilde{M}-1),
\ee
where $x_{t,0},\cdots,x_{t,\tilde{M}-1}$ are the known transmitted symbols in the training frame. Similarly, the last equation becomes
\be
\label{eqnNE:eq} y(n+\tilde{M}-N_{\mathrm{E}})=\sum_{l=0}^{M+L-1} x_{t,\tilde{M}-N_{\mathrm{E}}-l} \tilde{h}_l + z(n+\tilde{M}-N_{\mathrm{E}}).
\ee

In~(\ref{eqnNE:eq}) when $l=M+L-1$, we have $x_{t,\tilde{M}-N_{\mathrm{E}}-l}=x_{t,\tilde{M}-N_{\mathrm{E}}-M-L+1}$. Therefore, in order to guarantee that $x_{t,\tilde{M}-N_{\mathrm{E}}-l}$ is a valid training symbol the following needs to be satisfied:
\be
\tilde{M}-M-L-N_{\mathrm{E}}+1 \geq 0.
\ee
Therefore, $\tilde{M}\geq M+L+N_{\mathrm{E}}-1$. Choosing $\tilde{M}=M+L+N_{\mathrm{E}}-1$ (to reduce the training overhead) and stacking $N_{\mathrm{E}}$ received samples in a column vector we get
\be
\label{sys:eq}
\yb=\tilde{\Xb}_t \tilde{\hb} +\zb,
\ee
 where
 \be
 \yb=\left [y(n+\tilde{M}-1), y(n+\tilde{M}-2), \cdots, y(n+\tilde{M}-N_{\mathrm{E}})   \right ] ^{\sT}
 \ee
 is the known received vector of size $N_{\mathrm{E}}$,
 \be
 \tilde{\Xb}_t=\left [ \begin{array}{cccc}
                         x_{t,\tilde{M}-1} & x_{t,\tilde{M}-2} & \cdots & x_{t,N_{\mathrm{E}}-1} \\
                         x_{t,\tilde{M}-2} &  x_{t,\tilde{M}-3} &  & \vdots \\
                         \vdots &  & \ddots &  \\
                         x_{t,\tilde{M}-N_{\mathrm{E}}} & \cdots &  & x_{t,0}
                       \end{array}
 \right]
 \ee
is the $N_{\mathrm{E}} \times (M+L) $ measurement matrix constructed from known training symbols, and $\zb$ is the noise vector
 \be
 \zb\!=\!\!\left [z(n+\tilde{M}-1), z(n+\tilde{M}-2), \cdots, z(n+\tilde{M}-N_{\mathrm{E}})   \right ] ^{\sT}.
 \ee
  \subsection{Classical Solution}
 \label{classical:sec}
The system of equations in (\ref{sys:eq}) can be solved to estimate the combined CIR vector $\tilde{\hb}$. The classical solution of this problem (which does not exploit the sparsity of $\tilde{\hb}$) is given by
\be
\label{jointclassical:eq}
\hat{\tilde{\hb}}_{\mathrm{classical}}=\tilde{\Xb}_t^{\dagger}\yb,
\ee
where $\tilde{\Xb}_t^{\dagger}$ is the pseudo-inverse of $\tilde{\Xb}_t$ and this solution is called the  least-squares solution if the measurement matrix $\tilde{\Xb}_t$ is a tall matrix and it is called the minimum-norm solution if $\tilde{\Xb}_t$ is a fat matrix. The problems with this classical solution are as follows:
\begin{enumerate}
\item To obtain an accurate estimate of $\tilde{\hb}$, the number of equations $N_{\mathrm{E}}$ may be prohibitively large which increases the computational complexity.
\item The solution is not guaranteed to be a sparse solution although we know that $\tilde{\hb}$ is a sparse vector.
\end{enumerate}

\subsection{Proposed Sparsity-Aware Approach}
\label{omp:sec}
Since $\tilde{\hb}$ is a sparse vector, the problem in~(\ref{sys:eq}) can be viewed as a sparse signal recovery problem which can be formulated as follows
\be
\label{sparse:eq}
\underset{\tilde{\hb} \in \mathbb {C}^{M+L}}{\arg\min} \Vert \tilde{\hb}\Vert_0,~  \mathrm{s.t.} \: \Vert\yb- \tilde{\Xb}_t \tilde{\hb}\Vert_2^2 \leq \epsilon,
\ee
where $\Vert \tilde{\hb}\Vert_0$ is the number of non-zero elements in $\tilde{\hb}$. The notation $\Vert \centerdot \Vert_2 $ denotes the $\ell_2$-norm and $\epsilon$ can be chosen as a function of the noise variance.

To reduce the computational complexity of the Np-hard problem in~(\ref{sparse:eq}), the $\ell_0$-norm minimization is relaxed to an $\ell_1$-norm minimization problem and solved using the computationally efficient orthogonal matching pursuit (OMP) method. This solution is denoted by
\be
\label{sparseOMP:eq}
\hat{\tilde{\hb}}_{\mathrm{omp}}=\mathrm{OMP} (\yb, \tilde{\Xb}_t, \mathrm{stopping\:criterion}),
\ee
where the stopping criterion can be a predefined sparsity level on $ \Vert \tilde{\hb}\Vert_0$.

\subsection{Conventional Approach}
\label{conv:sec}
The conventional approach for frame synchronization and channel estimation is performing them separately. First, by cross-correlating the received data symbols with the training symbols, an estimate of the frame boundary, $\hat{\bar{D}}$ in $[0, M-1]$ is obtained\footnote{The cross-correlation method is optimal for frequency-flat channels and becomes suboptimal for multipath frequency-selective channels.}. Then, the equations in~(\ref{sys:eq}) can be expressed compactly as follows
\be
\label{sys2:eq}
\yb=\Xb_t \hb +\zb,
\ee
 where
 \be
\Xb_t=\left [ \begin{array}{cccc}
                         x_{t,\tilde{M}-\hat{\bar{D}}-1} & x_{t,\tilde{M}-\hat{\bar{D}}-2} & \cdots & x_{t,\tilde{M}-\hat{\bar{D}}-L-1} \\
                         x_{t,\tilde{M}-\hat{\bar{D}}-2} &  x_{t,\tilde{M}-\hat{\bar{D}}-3} &  & \vdots \\
                         \vdots &  & \ddots &  \\
                         x_{t,\tilde{M}-\hat{\bar{D}}-N_{\mathrm{E}}} & \cdots &  & x_{t,\tilde{M}-\hat{\bar{D}}-L-N_{\mathrm{E}}}
                       \end{array}
 \right]
 \ee
is the $N_{\mathrm{E}} \times (L+1) $ measurement matrix constructed from known training symbols.

To gain some insight from~(\ref{sys2:eq}), the following observations are in order:
\begin{enumerate}
  \item The first equation defined by the first row of $\Xb_t$ is given in~(\ref{first:eq}) where $\bar{D}$ is replaced by $\hat{\bar{D}}$.
  \item  $\Xb_t$ is a sub-matrix of $\tilde{\Xb}_t$. For example if $\hat{\bar{D}}=0$, then $\Xb_t$ consists of the first $L+1$ columns of $\tilde{\Xb}_t$. If $\hat{\bar{D}}=M-1$, then $\Xb_t$ consists of the last $L+1$ columns of $\tilde{\Xb}_t$.
  \item Unlike $\tilde{\hb}$ which will always be a sparse vector, $\hb$ may or may not be a sparse vector.
\end{enumerate}

If $\hb$ is not sparse, then the classical least-squares solution $\hat{\hb}=\Xb_t^{\dagger}\yb$
is a reasonable low complexity approach to solve this problem although sparse methods can still be applied if $\hb$ is a sparse CIR vector. Once $\hat{\hb}$ is obtained, the combined CIR vector estimate becomes
\be
\hat{\tilde{\hb}}_{\mathrm{conv}}= [\underbrace{0, 0, \cdots, 0}_{\hat{\bar{D}} \: \mathrm{zeros}} \hat{\hb}^{\sT}, \underbrace{0, 0, \cdots, 0}_{M-\hat{\bar{D}}-1 \: \mathrm{zeros}}  ]^{\sT}.
\ee

\section{Equalizer Design Based on Channel Estimates}
\label{equalizer:sec}
\begin{figure}[!t]
\centering
\includegraphics[width=3 in]{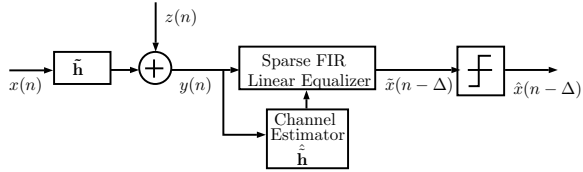}
\caption{The complete system model illustrating the combined channel, channel estimator and the equalizer.}
\label{comsys:fig}
\end{figure}

The multi-path channel introduces inter-symbol interference which should be equalized before the decisions on the symbols are made. The complete system model is illutrated in Fig.~\ref{comsys:fig}.
Here, the channel estimate from the previous section is used to design the equalizer. The performance of the equalizer will depend on the accuracy of the channel estimate. Therefore, the performance of different channel estimators from the previous section can be compared by evaluating the equalizer performance. To reduce the complexity of the equalizer, we implement the sparse FIR linear equalizer design of~\cite{abbasi&globalsip:2015,abbasi&twc:17}. The output of the equalizer is the soft estimate of the transmitted symbol $x(n)$ and can be expressed as $\tilde{x}(n)=\sum_{k=0}^{N-1} y(n-k)w_k,$
where $w_k$ is the $k$-th element of the equalizer vector $\wb$ and $N$ is the length of the equalizer. The performance metric adopted is the MSE defined as follows
\be
\mathrm{MSE}=\Expect{|\tilde{x}(n)-x(n-\Delta)|^2},
\ee
where $\Delta$ is the equalizer delay which is optimized to reduce the MSE. Finally, the transmitted symbols' estimates are calculated using a decision device based on the type of signal constellation used.

\section{Simulation and Experimental Results}

\subsection{Simulation Results}

\begin{figure}[t]
\centering
\includegraphics[width=2.5 in]{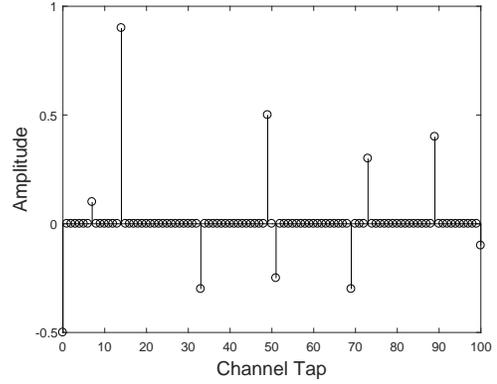}
\caption{The CIR $\hb$ used in the simulations.}
\label{channel:fig}
\end{figure}

The CIR that we assume in the simulation is similar to the channel in~\cite{cotter&tc:2002} and is shown in Fig.~\ref{channel:fig}.
\begin{figure*}[!h]
\centering
\includegraphics[width=7 in]{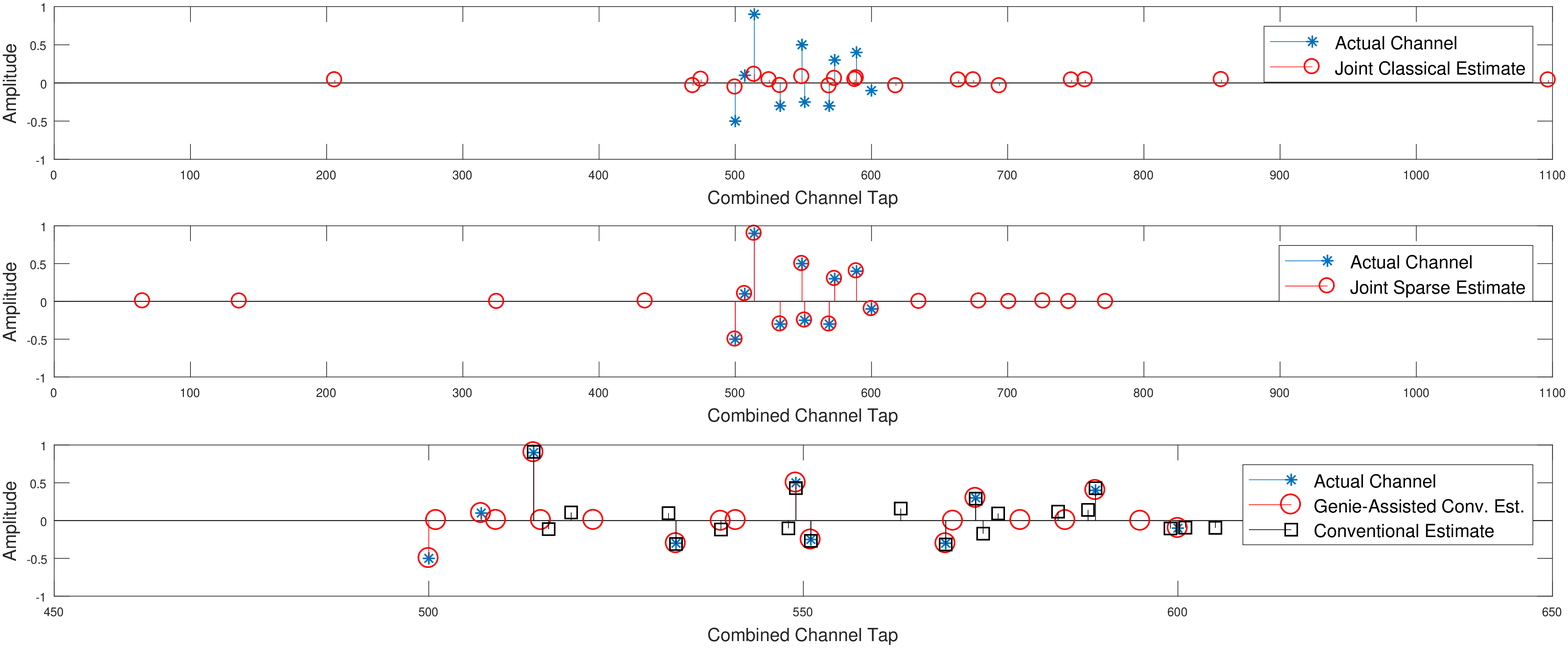}
\caption{The channel estimates obtained from classical method, joint sparse method, and conventional method with and without genie assistance. The last subplot showing conventional method has been zoomed in so that no estimate remains outside of the plot for better viewing.}
\label{channelestimate:fig}
\end{figure*}
This CIR has $L+1=101$ taps with 10 non-zero taps given by: $h_0=-0.5$, $h_7=0.1$, $h_{14}=0.9$, $h_{33}=-0.3$, $h_{49}=0.5$, $h_{51}=-0.25$, $h_{69}=-0.3$, $h_{73}=0.3$, $h_{89}=0.4$, and $h_{100}=-0.1$. The data frame length is $M=1000$. For channel estimation, $N_{\mathrm{E}}=148$ equations are used. The length of the training frame is $\tilde{M}=M+L+N_{\mathrm{E}}-1=1247$. We set the frame boundary $\bar{D}=500$.

We assume that we do not know the number of non-zero taps in the CIR and we can only have an upper bound on it which we set as 20. For the conventional approach, the least squares solution is used. The channel estimates obtained using the methods introduced in Section~\ref{joint:sec} are shown in Fig.~\ref{channelestimate:fig} where BPSK modulation is used and SNR is set to 20 dB.

For the conventional method, we also show the genie-assisted results where the receiver knows the correct frame boundary $\bar{D}=500$. Otherwise, when $\hat{\bar{D}}$ is obtained using cross-correlation of the received data with the transmitted training frame, $\hat{\bar{D}}=514$ is obtained which corresponds to the location of the strongest multi-path component. By visual inspection of the estimates, we see that the classical approach performs poorly compared to all other approaches. The joint sparse estimate and the genie-assisted conventional method have similar performance. Note that the conventional method without genie assistance is unable to estimate the 2 taps before the strongest tap at time index 514.

\begin{figure}[!h]
\centering
\includegraphics[width=2.8 in]{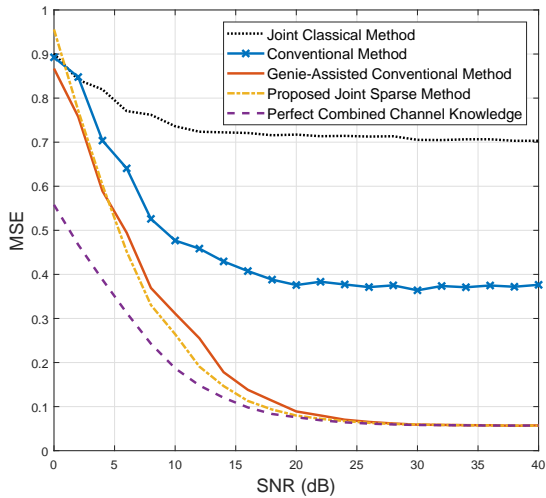}
\caption{Equalizer MSE performance of different methods using simulations.}
\label{sim1:fig}
\end{figure}

 Using the channel estimates from different methods and the ideal channel to design the equalizer, we illustrate the MSE performance in Fig.~\ref{sim1:fig}. The sparse FIR equalizer based on~\cite{abbasi&globalsip:2015} has 1200 taps with only 200 active taps. We observe that the proposed joint sparse method outperforms the genie-assisted conventional method for SNR larger than 5 dB. Furthermore, for SNR greater than 20 dB, the joint sparse method achieves same performance as in perfect channel knowledge with perfect frame synchronization. The performance of the conventional method is not satisfactory due to the errors in identifying the frame boundary.

\subsection{Experimental Results with USRPs}

\begin{figure}[!t]
\centering
\includegraphics[width=2.7in]{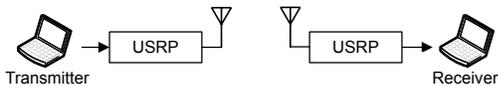}
\caption{System model for USRP experiments}
\label{model:fig}
\end{figure}

The performance of the proposed method was also evaluated using our wireless testbed at Qatar University consisting of a transmitter and a receiver where the transmitter and the receiver are composed of one laptop computer connected to a USRP N210 device~\cite{ettus} as shown in Fig.~\ref{model:fig}. Baseband processing of the signals as well as the communication with the USRP devices are done in MATLAB.

\begin{figure*}[!h]
\centering
\includegraphics[width=5.3 in]{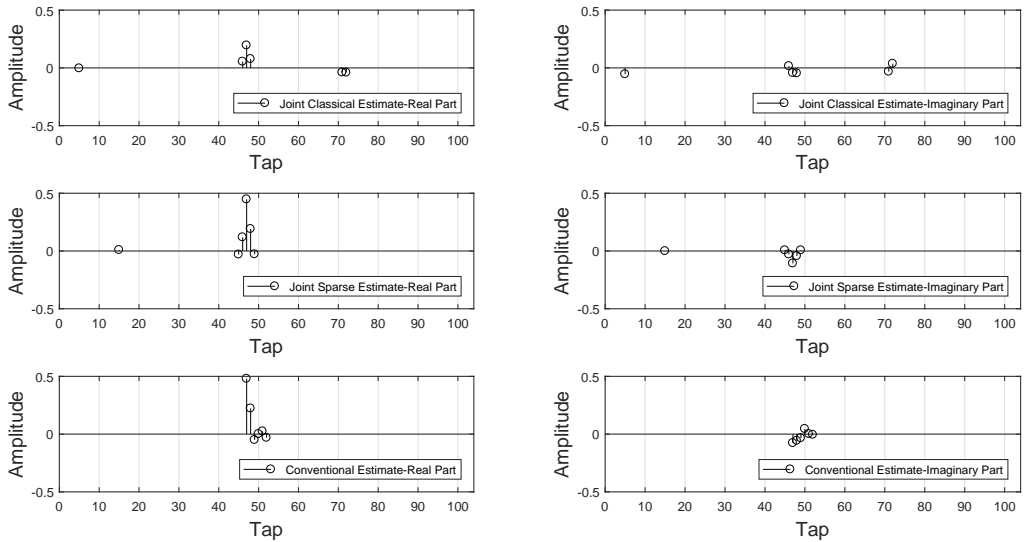}
\caption{The channel estimates obtained from classical method, joint sparse method, and conventional method with and without genie assistance.  }
\label{channelestimates2:fig}
\end{figure*}

The testbed is based on QPSK transmitter~\cite{qpsk1} and receiver~\cite{qpsk2} examples in MATLAB. We select a training frame period $P=10$. Data frames contain $M=100$ QPSK symbols whereas the training frame contains $\tilde{M}=147$ QPSK symbols. The symbols are up-sampled by 4 and passed through a raised cosine transmit filter. These samples are transmitted in the 2.4 GHz ISM band at a sampling rate $f_{\mathrm{s}}=200$~kHz. The same sampling rate is used at the receiver as well. The received samples go through automatic gain control (AGC) and a raised cosine receive filter. At the output of this filter, the over-sampling factor is reduced to 2. After coarse frequency compensation and fine frequency compensation, timing recovery is performed. During timing recovery over-sampling is reduced to 1 and the receiver clock is synchronized to the transmitter clock. However, the frame boundary is not determined yet. The channel needs to be equalized as well. At this point we can apply the different methods presented in this paper for frame synchronization and channel estimation.

In our USRP-based experiments, the multi-path symbol-spaced CIR used to generate the received samples was $\hb=[1~0.7]^{\sT}$. Fig.~\ref{channelestimates2:fig} shows the channel estimates using $N_{\mathrm{E}}=43$ equations where the real and imaginary parts of the channel estimates are drawn seperately. The channel is assumed to contain $L+1=6$ non-zero taps. At the top of the figure we have the joint classical method where the channel is estimated using~(\ref{jointclassical:eq}). Next, the channel estimate obtained using the proposed method is shown where the sparse OMP algorithm in~(\ref{sparseOMP:eq}) is used for channel estimation. The conventional method where the frame synchronization and channel estimation are performed separately, is at the bottom of Fig.~\ref{channelestimates2:fig}. We note that both methods seem to locate the frame boundary; however, it is obvious that the classical method performs the worst as its estimate is spread across the frame.

\begin{figure}[!h]
\centering
\includegraphics[width=2.8 in]{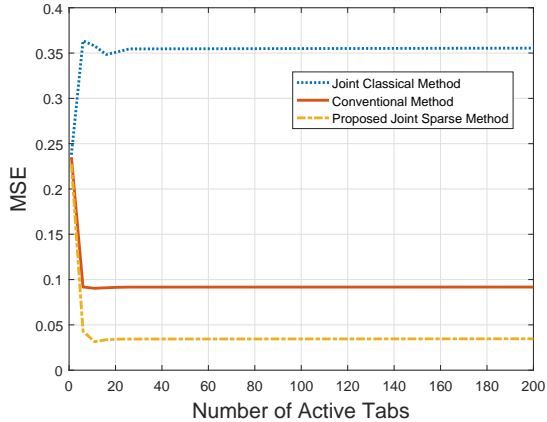}
\caption{MSE performance of different methods as a function of the number of active equalization taps.}
\label{sim2:fig}
\end{figure}

To assess the performance of each method, we perform the sparse FIR linear equalization of~\cite{abbasi&globalsip:2015} as explained in Section~\ref{equalizer:sec}. The equalization is performed for data frames using the channel estimates obtained during the training frame. Fig.~\ref{sim2:fig} shows the MSE results as a function of the number of active taps used in the equalizer out of a total of 200 taps. Note that the proposed method outperforms the conventional and the joint classical methods. Because the joint classical method results in an inaccurate channel estimate, the MSE increases as the number of active taps increases. We also note that around 10 active taps are sufficient for the equalizer to converge to the optimum performance.

\section{Conclusion}

In this paper, we proposed a joint frame synchronization and channel estimation method for frame-based communication systems. The proposed sparsity-aware method attains superior performance compared to the conventional method where frame synchronization and channel estimation are done separately. A full frame is dedicated to training; however, the overhead due to training can be minimized by increasing the period of the training frame by taking the coherence time of the channel into consideration.  Future research includes extensions to multi-antenna and multi-user systems and investigating more sophisticated sparse recovery algorithms than the simple OMP considered in this paper.

\bibliographystyle{IEEEtran}

\bibliography{draft}

\end{document}